\documentclass[preprint2]{aastex6}
\pdfoutput=1 
\usepackage{amsmath,amstext}
\usepackage{booktabs}
\usepackage{xspace}
\usepackage{hyperref}
\usepackage[T1]{fontenc}
\usepackage{lipsum}

\usepackage{natbib}
\usepackage{graphicx} 

\usepackage{enumitem}

\accepted{2017 September 1}
\published{2017 October 6}

\shorttitle{Radio lobe dynamics in NGC 4472}
\shortauthors{M. Gendron-Marsolais et al.}

\begin{document}

\title{Uplift, feedback and buoyancy: radio lobe dynamics in NGC 4472}


\author{
M. Gendron-Marsolais\altaffilmark{1,2},
R. P. Kraft\altaffilmark{2},
A. Bogdan\altaffilmark{2},
J. Hlavacek-Larrondo\altaffilmark{1,2},
W. R. Forman\altaffilmark{2},
C. Jones\altaffilmark{2},
Y. Su\altaffilmark{2},
P. Nulsen\altaffilmark{2}, 
S. W. Randall\altaffilmark{2}
and E. Roediger\altaffilmark{3}}

\altaffiltext{1}{Departement de Physique, Universite de Montreal, Montreal, QC H3C 3J7, Canada}
\altaffiltext{2}{Harvard-Smithsonian Center for Astrophysics, 60 Garden Street, Cambridge, MA 02138, USA}
\altaffiltext{3}{E.A. Milne Centre for Astrophysics, Department of Physics and Mathematics, University of Hull, Hull, HU6 7RX, UK}

\begin{abstract}
We present results from deep (380 ks) \textit{Chandra} observations of the AGN outburst in the
massive early-type galaxy NGC 4472.
We detect cavities in the gas coincident with the radio lobes and estimate the eastern and western lobe enthalpy to be $(1.1 \pm 0.5 )\times 10^{56}$ erg and $(3 \pm 1 )\times 10^{56}$ erg, and the average power required to inflate the lobes to be $(1.8 \pm 0.9)\times 10^{41}$ erg s$^{-1}$ and $(6 \pm 3)\times 10^{41}$ erg s$^{-1}$, respectively.  We also detect enhanced X-ray rims around the radio lobes with sharp surface brightness discontinuities between the shells and the ambient gas. The temperature of the gas in the shells is less than that of the ambient medium, suggesting that they are not AGN-driven shocks but rather gas uplifted from the core by the buoyant rise of the radio bubbles. 
We estimate the energy required to lift the gas to be up to $(1.1 \pm 0.3 )\times 10^{56}$ erg and $(3 \pm 1 )\times 10^{56}$ erg for the eastern and western rim respectively, constituting a significant fraction of the total outburst energy. A more conservative estimate suggests that the gas in the rim was uplifted a smaller distance, requiring only $20-25\%$ of this energy. In either case, if a significant fraction of this uplift energy is thermalized via hydrodynamic instabilities or thermal conduction, our results suggest that it could be an important source of heating in cool core clusters and groups.
We also find evidence for a central abundance drop in NGC 4472. 
The iron abundance profile shows that the region along the cavity system has a lower metallicity than the surrounding, undisturbed gas, similar to the central region. This also shows that  bubbles have lifted low-metallicity gas from the center.
\end{abstract}


\section{Introduction}

\begin{figure*}
    \centering
    \includegraphics[width=0.48\textwidth]{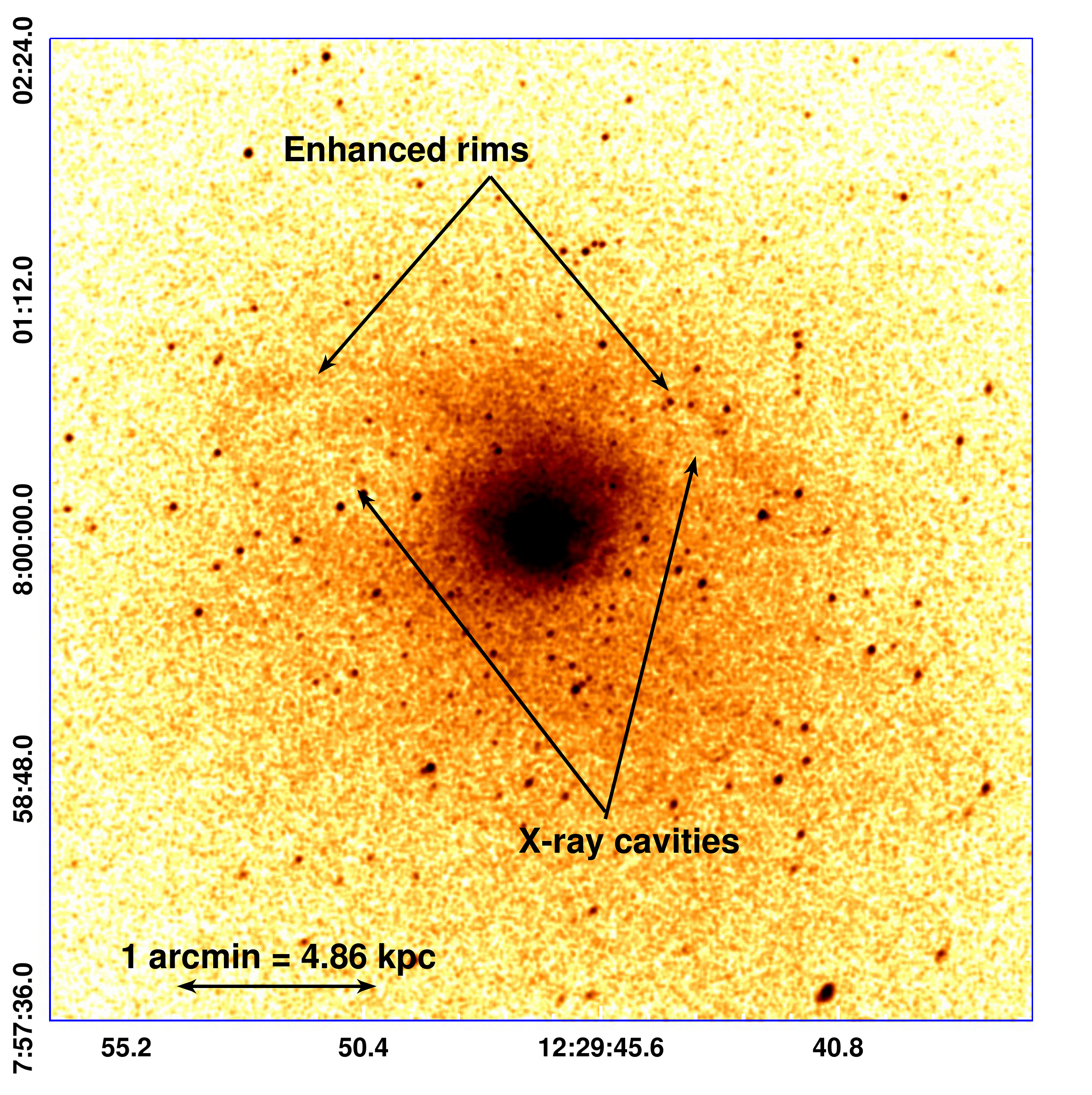}
    \includegraphics[width=0.48\textwidth]{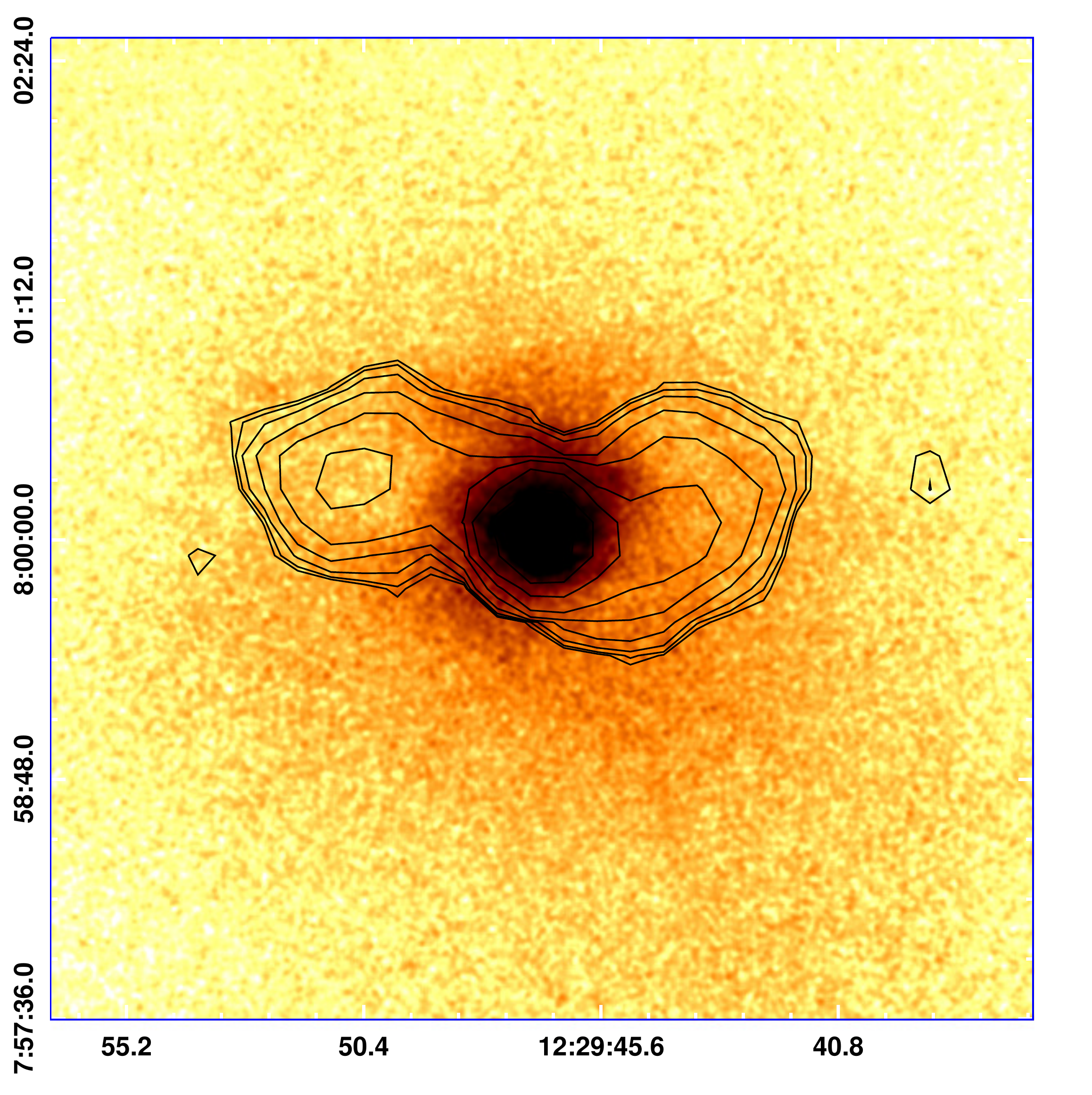}
\caption{Left - Raw image of the 369.60 ks merged \textit{Chandra} observations in the 0.5-2 keV band, smoothed with a 2 pixel Gaussian function. The main X-ray structures are identified: two X-ray cavities, each surrounded by enhanced rims of X-ray emission. Right - Background-subtracted and exposure-corrected image in the 0.5-2 keV band, smoothed with a 3 pixel Gaussian function, point sources removed and replaced by regions interpolated from the surrounding background using \textsc{dmfilth}, with L-band (20 cm) VLA radio contours starting at $ 3 \sigma = 0.9$ mJy/beam.}\label{fig:ngc4472_merged_500-2000ev_raw_and_dmfilth}
\end{figure*}

 \newcommand{\rr}{\raggedright}
  \newcommand{\tn}{\tabularnewline}
  
  { \renewcommand{\arraystretch}{1.2}
\begin{table*}
\centering
 \caption{\textit{Chandra} observations}
 \label{tab:obsid}
 \begin{tabular}{  p{1.3cm}  | p{2.5cm}  p{2.5cm}  p{1.8cm} p{1.8cm} p{1.8cm} }
  \hline
 \rr  Obs ID & \rr Total exposure time (ks) & \rr Cleaned exposure time (ks) & \rr Data mode & \rr Number of counts \textsuperscript{*} & \rr Start date \tn \hline
  321     & 39.59   & 35.84   & VFAINT & 103330  & 2000-06-12 \\
  11274 & 39.67   & 39.67   &  FAINT  & 105240  & 2010-02-27 \\
  12888 & 159.31 & 159.31 & VFAINT & 403338 & 2011-02-21 \\
  12889 & 135.59 & 134.78 & VFAINT & 342530 & 2011-02-14 \\
  \hline
 \rr Total & 374.16 & 369.60 &   & 954438 &  \\
 \hline
\multicolumn{4}{l}{\textsuperscript{*}\footnotesize{Number of counts in chip S3 in the broad band, from 0.5-7 keV.}}
 \end{tabular}
\end{table*}
}

Active galactic nuclei (AGN) mechanical feedback is believed to compensate the radiative losses of the intraculster medium (ICM) in many clusters of galaxies (e.g. \citealt{churazov_asymmetric_2000,churazov_evolution_2001,birzan_systematic_2004,dunn_investigating_2006,rafferty_feedback-regulated_2006}).
Through this feedback process, energy's injected in the ICM via outflows produced from the accretion onto the central supermassive black hole (SMBH) of the dominant galaxy. 
Signatures of this process are directly observable at X-ray wavelengths as the inflation of jet-driven radio bubbles displaces the ICM and creates regions known as X-ray cavities (e.g. \citealt{bohringer_rosat_1993,mcnamara_chandra_2000,mcnamara_heating_2007,mcnamara_mechanical_2012}). 
The question of how the energy is transferred - through shocks, turbulent heating or sound waves - is still a matter of debate. One way to address this issue is to study individual nearby cases where the proximity of these objects gives the sensitivity and resolution needed to study AGN feedback in detail and provide a better understanding of this energy transfer.

AGN feedback has been well-studied in a large sample of galaxy groups and clusters 
(Perseus e.g. \cite{fabian_wide_2011}, Virgo e.g. \cite{forman_filaments_2007}, NGC 5813 e.g. \cite{randall_very_2015}, Centaurus e.g. \cite{sanders_very_2016} and distant clusters e.g. \citealt{hlavacek-larrondo_extreme_2012,hlavacek-larrondo_x-ray_2015}), but the underlying microphysics can only be studied in detail in the nearest objects. 

The early-type galaxy NGC 4472 (M49) is well suited for this type of study. It is the dominant member of a galaxy group, the brightest group galaxy (BGG), lying on the outskirts of the Virgo cluster (4 degrees south of M87). This massive  elliptical galaxy (stellar mass $\sim 8\times 10^{11} \text{ M}_{\odot}$, \citealt{cote_dynamics_2003}) is the most optically luminous galaxy in the local Universe (0.2 mag brighter than M87).

The hot X-ray corona ($\sim 1$ keV, $\text{L}_{\text{X}} \sim 10^{42}$ erg s$^{-1}$) in NGC 4472 was first observed by \cite{forman_hot_1985} with the \textit{Einstein Observatory}, whereas the first measurement of the abundance in the corona was accomplished with \textit{ROSAT} \citep{forman_heavy_1993}.
Further observations revealed an elongation of the X-ray emission in the northeast-southwest direction as well as a bow shock-like structure on the north side, both resulting from ram pressure stripping from the ICM in the Virgo cluster \citep{irwin_x-ray_1996}.
Several new structures - a symmetric bright central region, two X-ray cavities corresponding to the radio lobes, a drop in the surface brightness profile 14 kpc northeast from the nucleus and a tail of emission extending 37 kpc to the southwest from the nucleus - were found in the analysis of 40 ks of \textit{Chandra} ACIS-S observations on NGC 4472 \citep{biller_hot_2004}.
\cite{kraft_gas_2011} also presented a study of the hot gas dynamics in NGC 4472 based on 100 ks of \textit{XMM-Newton} observations. They found a surface brightness discontinuity at 21 kpc north of the nucleus, detected a $>60$ kpc long ram-pressure stripped tail to the southwest and two sets of cool filamentary arms (to the east and southwest, $\sim25$ kpc long) which were interpreted as gas uplifted by buoyantly rising radio bubbles. Those older bubbles are currently not observed at GHz radio frequencies.

NGC 4472 is therefore undergoing both internal (by the nuclear outburst) and external (by the interaction with the Virgo ICM) disturbances. 
In this paper, we study the central regions of the galaxy group, focusing on the effects of the mechanical AGN feedback on the X-ray gas.
In addition to the already existing $\sim 80$ ks, we present $\sim 300$ ks of new ACIS-S \textit{Chandra} observations (PI Kraft).

\textit{Chandra} observations and the data reduction are summarized in section \ref{Observations and data reduction}. In sections \ref{Analysis} and \ref{Abundance maps and profiles}, we analyze the structures found in the X-ray data and the metallicity distribution.
Section \ref{Discussion} discusses the results, comparing NGC 4472 to other groups and clusters. Finally, the results are summarized in section \ref{Conclusion}.

We assume a luminosity distance of 16.7 Mpc for NGC 4472 \citep{blakeslee_acs_2009} and an angular scale of 4.86 kpc arcmin$^{-1}$. This corresponds to a redshift of $z = 0.0038$, assuming $H_{0} = 69.6 \text{ km s}^{-1} \text{Mpc}^{-1}$, $\Omega_{\rm M} = 0.286$ and $\Omega_{\rm vac} = 0.714$. 
Unless mentioned otherwise, all uncertainties are one sigma.

\section{Observations and data reduction}\label{Observations and data reduction}

\subsection{\textit{Chandra} observations and data reduction}

We used new, deep \textit{Chandra} Advanced CCD Imaging Spectrometer (ACIS) observations (aimpoint on chip S3, OBSID 12888 and 12889, see table \ref{tab:obsid}) combined with archival data (OBSID 321 and 11274). The total exposure time is 374.16 ks and the merged dataset has about one million counts in chip S3 in the broad band from 0.5-7 keV. For our analysis, we used the X-ray-processing packages \textsc{ciao} version 4.7 and \textsc{xspec} version 12.9. 

After reprocessing the datasets using \textsc{chandra\_repro}, point sources were detected with \textsc{wavdetect} for each observation, both in the broad $0.5\,-\,7$ keV and in the hard $2\,-\,7$ keV bands. The default parameters of the \textsc{wavdetect} tool were used, except for the wavelet radii list in pixels (2, 4, 8, 16), the image and the PSF file of each observation were given as inputs.
Excluding the point sources found in the hard band, light curves were produced with \textsc{dmextract}. Flares were identified and removed with \textsc{deflare} and the \textsc{lc\_clean} routine, using default parameters (outliers more than $3\sigma$ from the mean count rate were removed). The exposure time remaining after the flare filtering is 369.60 ks.
The tool \textsc{merge\_obs} was used to merge all the observations together. \textsc{wavdetect} was used again on the merged images in the broad ($0.5\,-\,7$ keV) and the hard ($2\,-\,7$ keV) band. These point source lists were combined and a final list of 503 point sources was made after visual inspection of these regions. 
Using blank-sky background files tailored to each observation and normalized using the hard energy count rates ($9.5\,-\,12$ keV) to match them to each observation, a merged background image was also produced.
Finally, an exposure-corrected image was obtained dividing the merged image from the merged exposure map produced by the tool \textsc{merge\_obs}.

\subsection{Radio observations}

To constrain the shape of the X-ray cavities, we used an image in FITS format from the NRAO \textit{Very Large Array} (VLA) Archive Survey Images.
These images are produced from publicly available data processed with a pipeline in \textsc{AIPS}. 
In particular, we selected a 1.4 GHz (L-band, 20 cm) C-configuration image from observation AB0412 \citep{condon_radio_1988}. 
The image has a beam size of $18 \arcsec \times 18 \arcsec$ and a rms of $0.3 \text{ mJy}/\text{beam}$. Two radio lobes of about one arcmin - around 5 kpc - diameter are clearly visible in these observations, on either side of the AGN.
The flux density at 1.4 GHz of the nucleus is around 0.2 Jy and it is 8 mJy in the lobes.

\section{Analysis of the cavities and the enhanced X-ray emission rims}\label{Analysis}

The merged raw image of the X-ray observations is shown in Fig.~\ref{fig:ngc4472_merged_500-2000ev_raw_and_dmfilth}-left. Fig.~\ref{fig:ngc4472_merged_500-2000ev_raw_and_dmfilth}-right shows the background-subtracted, exposure-corrected, smoothed image where the point sources have been replaced by regions interpolated from the surrounding background using the tool \textsc{dmfilth}, with L-band (20 cm) VLA radio contours starting at $ 3 \sigma = 0.9$ mJy/beam.. As indicated by the arrows, clear cavities in the X-ray emission are detected at the position of the radio lobes, and rings of enhanced X-ray emission just beyond the lobes. 
To enhance deviations from the radial surface brightness profile, the merged point-source subtracted image was subtracted and divided by the radial average of the surface brightness profile estimated using 2.5 pixels thick annular regions centered on the X-ray peak of the image. This fractional residual image (see Fig.~\ref{fig:ngc4472_merged_dmfilth_500-2000ev_unsharpmask_reg_radio}-left) shows even more clearly the cavities' shapes as well as the enhanced rims enclosing them.
An analysis of these cavities and rims is presented in the following sections.

\subsection{Enhanced X-ray emission rims}
\label{Enhanced X-ray emission rims}

We explored two possible explanations for the presence of the enhanced X-ray emission rims around the cavities. First, if the bubbles are in a phase of supersonic expansion, these rims could be material pushed aside forming a shocked shell of gas around the bubbles (e.g. \citealt{fabian_deep_2003,forman_filaments_2007}). Second, if the bubbles are rising buoyantly, in a sub-sonic phase of expansion, then those could be gas uplifted from the center, dragged out by the rising bubbles (e.g. \citealt{fabian_chandra_2000,churazov_evolution_2001}).

\begin{figure*}
\centering 
   \includegraphics[width=0.38\textwidth]{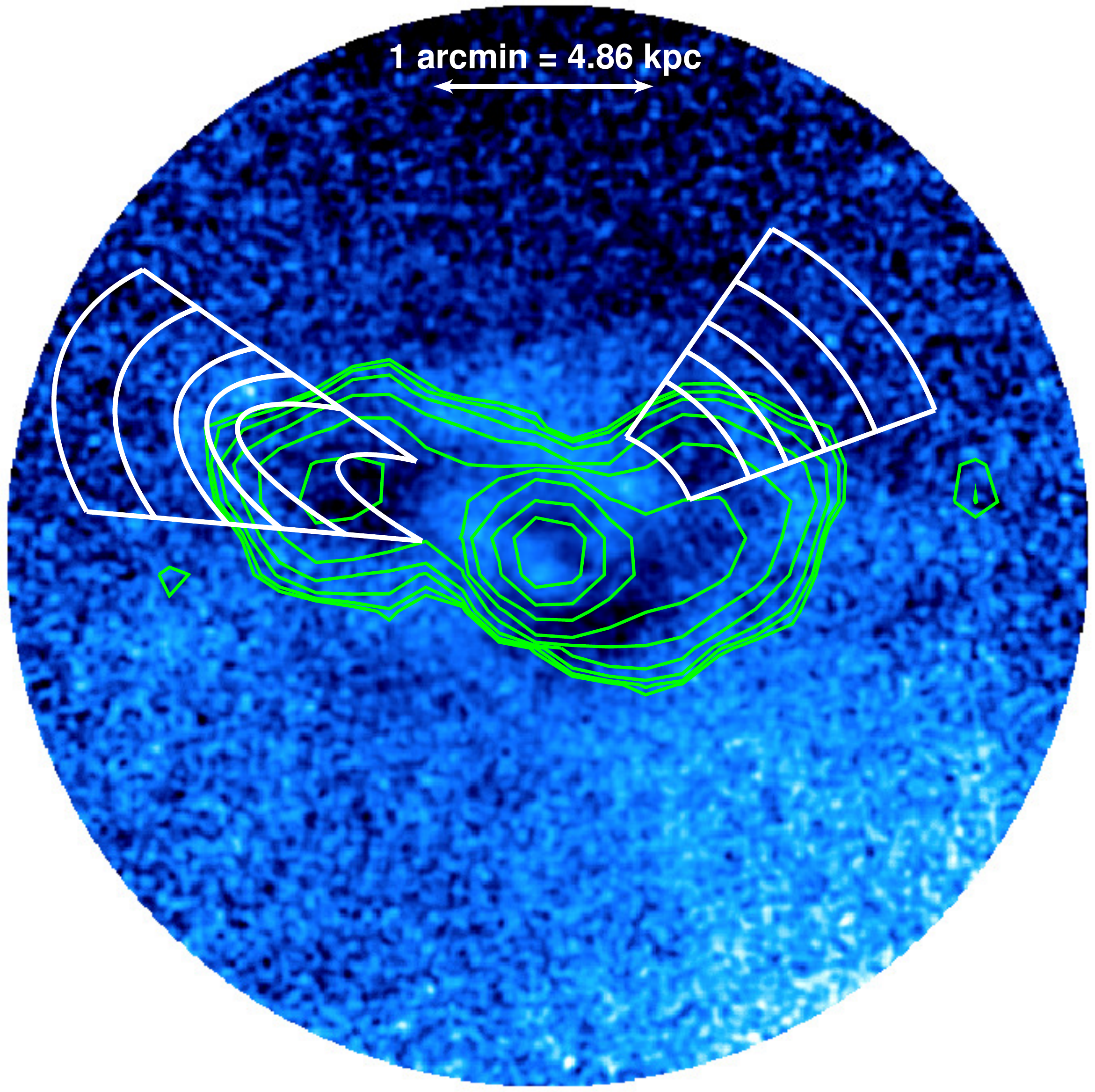}
   \includegraphics[width=0.5\textwidth]{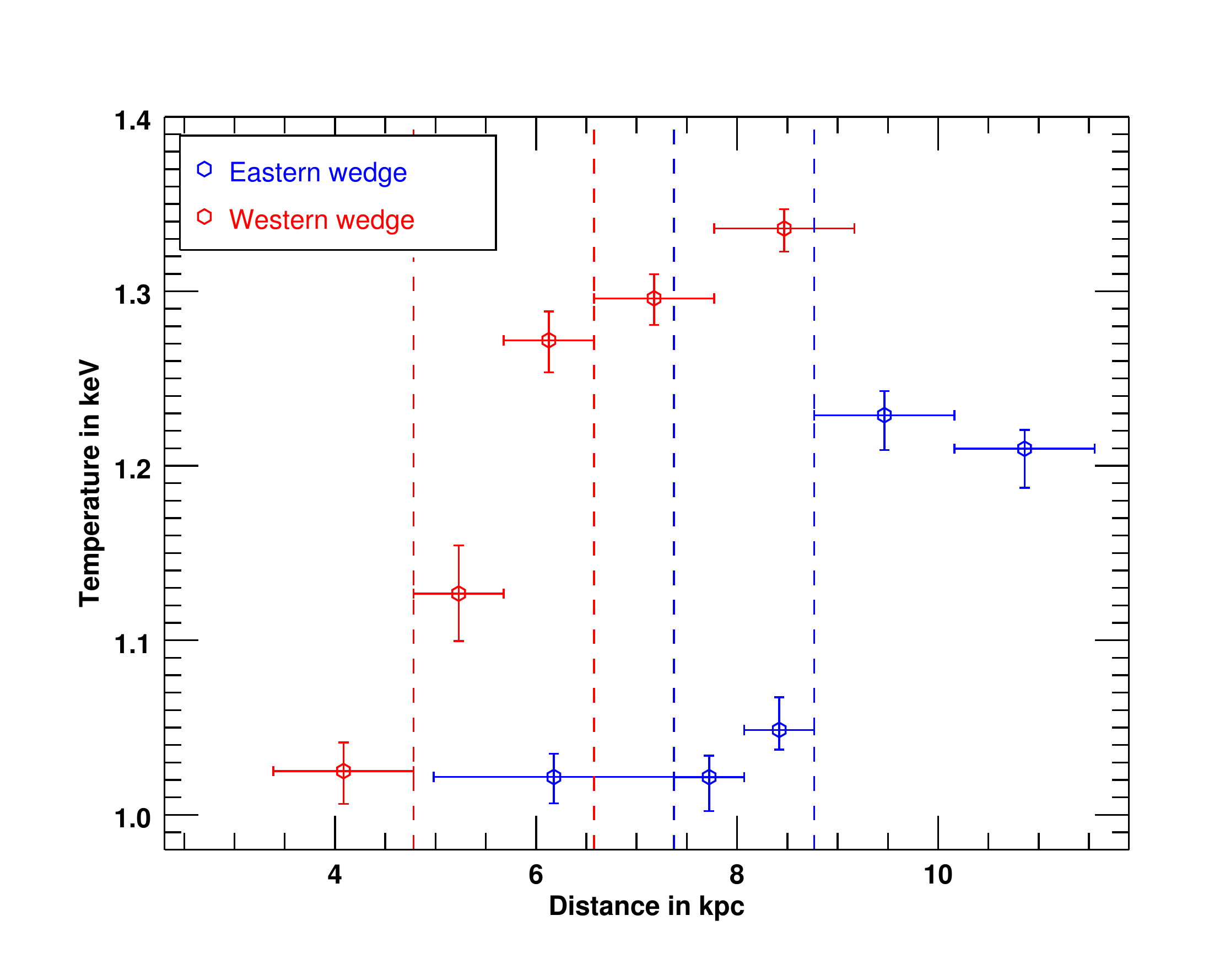}
\caption{\textbf{Left:} Fractional residual image in the 0.5-2 keV band, smoothed with a 2 pixel Gaussian function. L-band (20 cm) VLA radio contours starting at $ 3 \sigma = 0.9$ mJy/beam are shown in green. Regions across each rim, following closely the shapes of the edges, are shown in white. \textbf{Right:} Projected temperature profile across each rim. The red points are for the western wedge and the blue points are for the eastern wedge. The distances on the horizontal axis are the distances from the X-ray center to the center of each region. The dashed lines show the position of the inner and outer bounds of the two rims.}
\label{fig:ngc4472_merged_dmfilth_500-2000ev_unsharpmask_reg_radio}  
\label{fig:temp_profile}
\end{figure*}

To distinguish between the two alternatives, we created a temperature profile across each rim. 
Using the fractional residual image, five semi-elliptical annuli were closely matched to the shape of each rim (see Fig.~\ref{fig:ngc4472_merged_dmfilth_500-2000ev_unsharpmask_reg_radio}). The first region lies inside the cavity, two regions lie inside the shell and two other regions lie outside the rim. 
The spectrum in each of those regions was extracted from each observation individually with \textsc{specextract} and fitted simultaneously with \textsc{xspec} in the $0.4-2.5$ keV band fixing the absorption by foreground gas in our Galaxy\footnote{The neutral hydrogen column density was calculated using the COLDEN tool: \url{http://cxc.harvard.edu/toolkit/colden.jsp} and was the same value used in \cite{kraft_gas_2011}} to $1.62 \times 10^{20} \text{ cm}^{-2}$. 
Considering the high number of counts and the low temperature of NGC 4472, we used a \textsc{phabs$\ast$vapec} model, allowing the O, Ne, Mg, Si and Fe abundances to vary and fixing the others (C, N, Al, S, Ar, Ca, Ni) to 0.3.
The abundance table used for the modelling of plasma emission and photoelectric absorption in all \textsc{xspec} models used in this paper were taking from \cite{anders_abundances_1989}.
We used the atomic database AtomDB version 2.0.2.
The temperature profile across each rim is shown in Fig.~\ref{fig:temp_profile}.
For both rims, the temperature inside the rim ($1.02_{-0.02}^{+0.01}$ and $1.05_{-0.01}^{+0.02}$ for the eastern rim, $1.13 \pm 0.03$ and $1.27 \pm 0.02$ for the western rim) is lower than outside it ($1.23_{-0.02}^{+0.01}$ for the eastern rim and $1.30_{-0.02}^{+0.01}$ for the western rim).
The rims are colder than the exterior gas, suggesting that the observed surface brightness jumps are not associated with AGN-driven shocks, but rather with cold gas being dragged out from the center by buoyantly rising bubbles. Similar $\sim25$ kpc long filamentary arms have been found in XMM-Newton observations starting at a radius of $\sim 15$ kpc \citep{kraft_gas_2011} and could also be gas uplifted by older detached bubbles. 
Realistic numerical simulations of AGN outbursts in clusters show similar shells with enhanced X-ray surface brightness around cavities (e.g. \citealt{heinz_answer_2006,morsony_swimming_2010}).
Numerical simulations of the evolution of buoyantly rising bubbles into the ICM also show cold gas being dragged out from the center in the wake of the bubbles (e.g. \citealt{churazov_evolution_2001,gardini_buoyant_2007,revaz_formation_2008}).


\begin{figure*}
\centering  
 \includegraphics[width=0.4\textwidth]{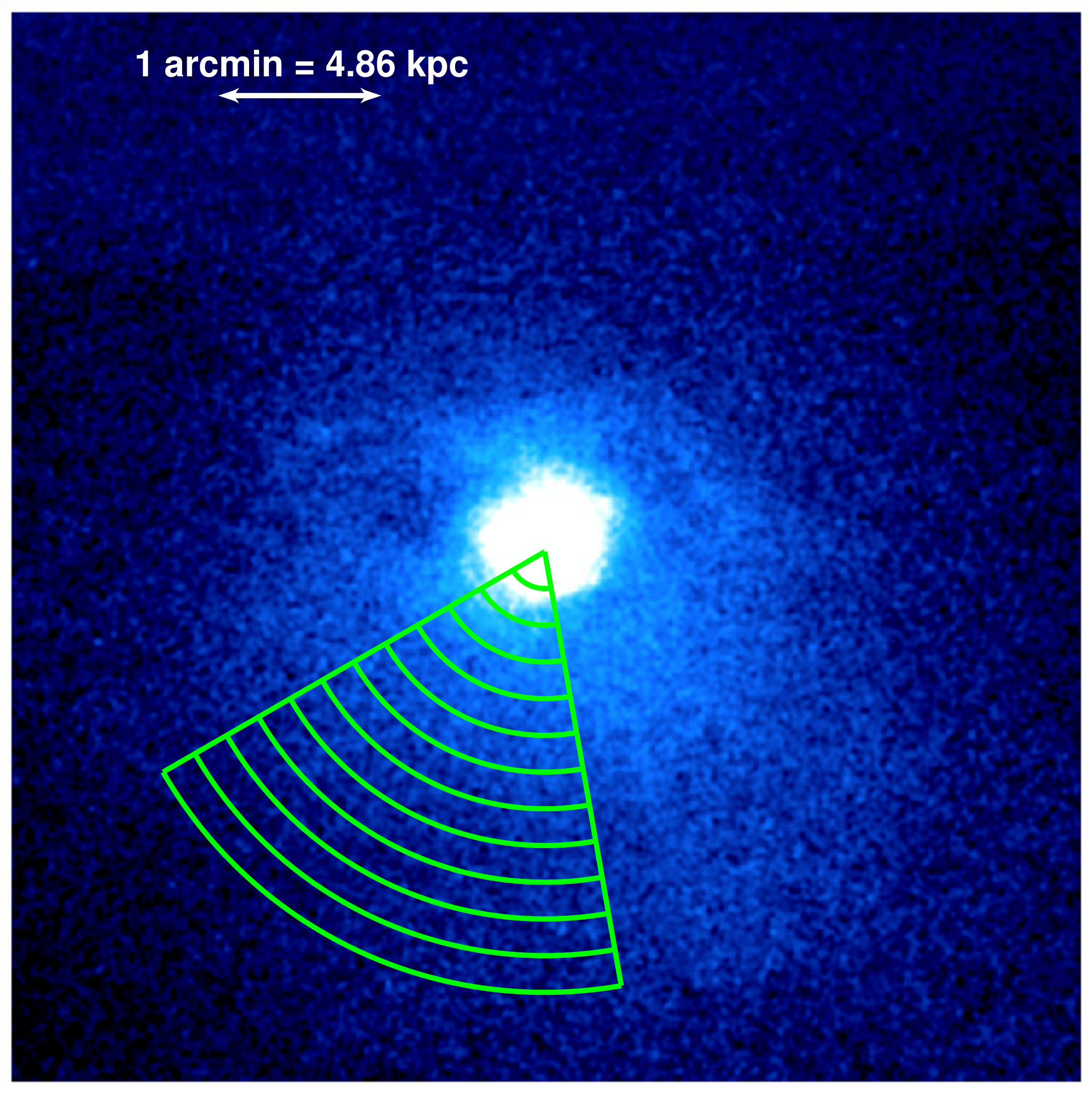}
\quad \quad 
 \includegraphics[width=0.45\textwidth]{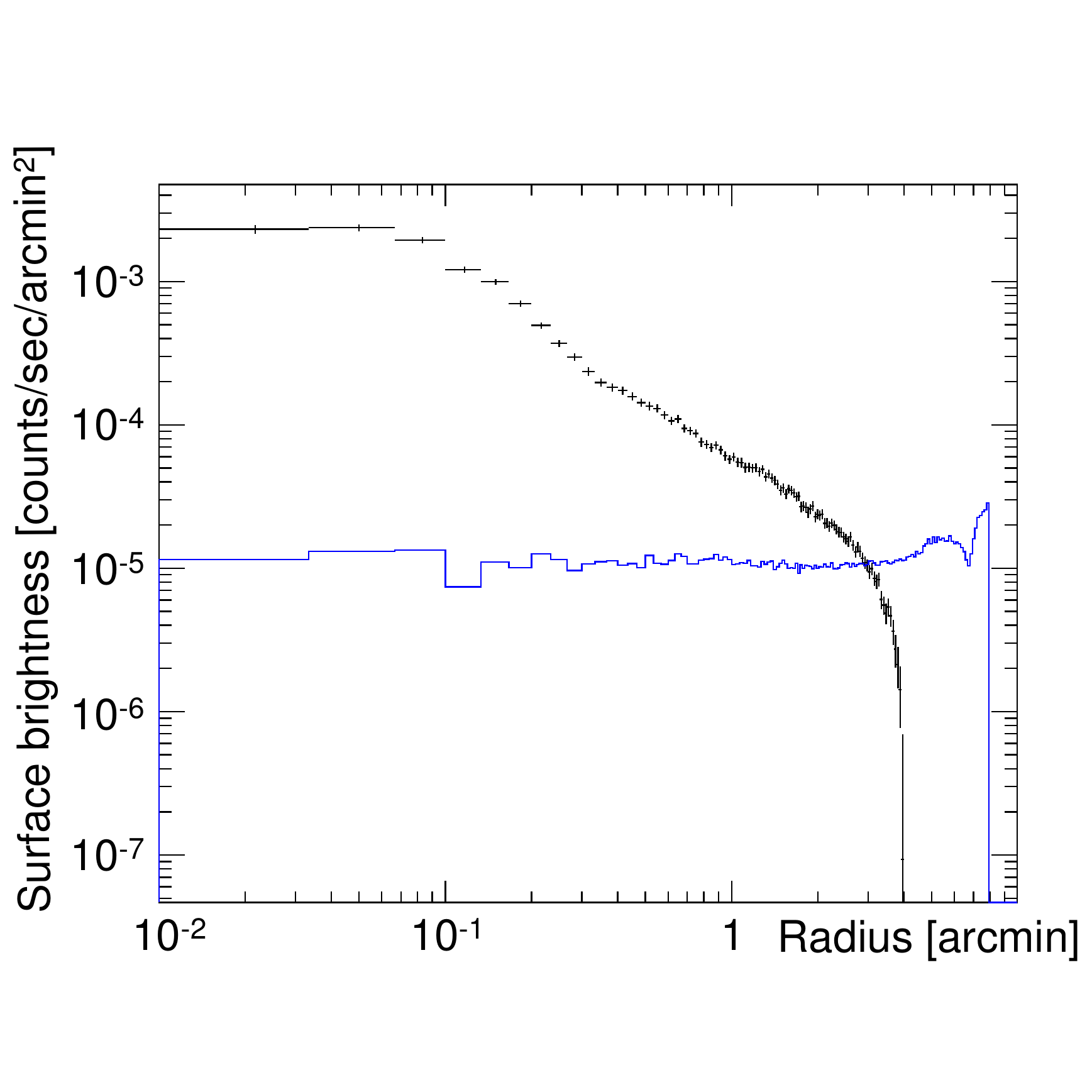}
 \caption{\textbf{Left:} Background-subtracted and exposure-corrected image smoothed with a 4 pixel Gaussian function, point sources removed, in the 0.5-2 keV band, with the 12 regions where the temperature, density and pressure profiles were extracted (see Fig.~\ref{fig:plot_profiles}). \textbf{Right:} Surface brightness profile (in black) from this wedge obtained with \textsc{Proffit}. The subtracted background from the merged blank-sky files is shown in blue.}
 \label{fig:ngc4472_merged_dmfilth_500-2000ev_bkgcorr_expcorr_smooth4_regtemp}
 \label{fig:surf_bri_profile_proffit}
\end{figure*}

\subsection{Outburst energy}

\begin{figure}
\hspace*{-1cm}
 \includegraphics[width=10cm]{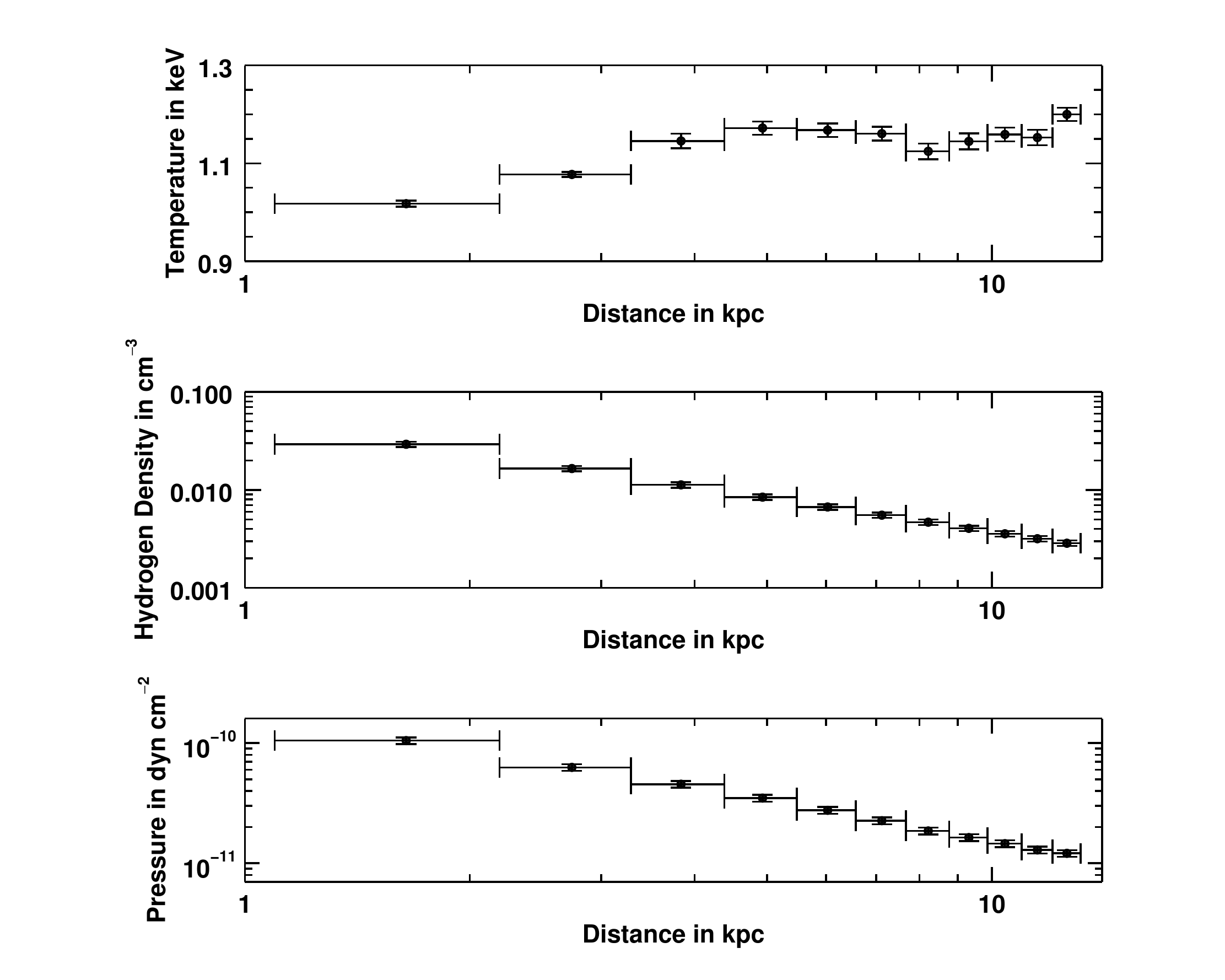}
 \caption{Projected temperature, density and pressure profiles of the southern wedge shown in Fig.~\ref{fig:ngc4472_merged_dmfilth_500-2000ev_bkgcorr_expcorr_smooth4_regtemp}.}
 \label{fig:plot_profiles}
\end{figure}

The presence of X-ray cavities allows the energy injected into the ICM by the jets to be measured.
To estimate the energetics of the lobes, we first need to build the density and temperature profiles.
Using \textsc{specextract}, we extracted spectra in 12 regions in the southern direction (see Fig.~\ref{fig:ngc4472_merged_dmfilth_500-2000ev_bkgcorr_expcorr_smooth4_regtemp}~-left). 
This wedge was chosen since it contains gas presumably unperturbed by the AGN outburst as there are no cavities in that direction. The northern direction seems to be more disturb with the presence of the edge of western rim. The same \textsc{phabs$\ast$vapec} model was fitted to the data with the previously described settings for the parameters, except that the Ne abundance was fixed to 0.3 to improve the fit.
We then used the \textsc{Proffit} software package \citep{eckert_cool-core_2011} to obtain a surface brightness profile for the wedge shown in Fig.~\ref{fig:surf_bri_profile_proffit}- left, providing the merged image, background and exposure map. We fitted this profile with a broken $\beta$-model projected along the line of sight:
\begin{eqnarray}
n_{\text{H}} (r) = 
\begin{cases}
n_{0}\left(1+ \left( \dfrac{r}{r_{c}} \right)^{2}   \right)^{-\dfrac{3\beta}{2}}  ,  0.27\arcmin< r < 1.25 \arcmin \\
n_{1} r^{-\alpha_{1}}  \; \; \quad \quad \quad \quad\quad\quad , 1.25\arcmin < r <2.75\arcmin
\end{cases}
\end{eqnarray}
where $r$ is the distance from the center in arcminutes. 
The best-fit model is shown in Fig.~\ref{fig:plot_profiles}. The best-fit parameters are $\beta=0.39 \pm 0.01$, $r_{c}=0.08\arcmin \pm 0.02\arcmin$ and $\alpha_{1}=1.29 \pm 0.03$. The density jump between the $\beta$-model and the power law is located $1.25\arcmin$ from the center and is small ($n_{0} / n_{1} = 0.86 \pm 0.03$). Indeed, the surface brightness discontinuity is hardly visible on the image.
Several different models were tested to fit the profile (such as a power-law, a $\beta$-model and a broken power-law model) as well as different fixed values for the cutting radius parameter, this broken $\beta$-model giving the lowest reduced $\chi$-squared value (0.377).

The density is obtained from the normalization provided by \textsc{xspec} and the total pressure is given by $P = 2.2 n_{\text{H}} k T$. The factor $2.2$ comes from the fact that we assume a total number density given by $n_{\text{total}}=n_e + n_{\text{H}}$ and a mean molecular weight of $\mu =0.61$, therefore giving $n_e = 1.2 \; n_{\text{H}}$.
The temperature, density and pressure profiles are shown in Fig.~\ref{fig:plot_profiles}.

We then calculated the enthalpy of the cavities, which is the sum of the work done to displace the gas plus the thermal energy inside the bubble. Considering a relativistic ideal gas (with a ratio of specific heats of $4/3$), the enthalpy is $4PV$. 
From the analysis of section \ref{Enhanced X-ray emission rims}, the radio bubbles appear to be rising buoyantly.
This means that their shape is more easily perturbed by the surrounding gas \citep{heinz_answer_2006} and also that Rayleigh-Taylor instabilities have modified the initially spherical bubbles into flattened mushroom shaped bubbles \citep{churazov_evolution_2001}. The main source of uncertainty in the calculation of the enthalpy is therefore the size of the lobes. We assumed reasonable minimum and maximum 3D shapes for each lobe, giving a range of possible volumes. 
Based on the residual fractional image and the radio emission, we assumed an ellipsoidal lobe for the shape of the eastern cavity and half of an ellipsoidal shell for the western cavity (Fig.~\ref{fig:ngc4472_merged_dmfilth_500-2000ev_unsharpmask_lobes_and_rims}),
giving $(8 \pm 4 )\times 10^{65} \text{ cm}^{3}$ and $(1.5 \pm 0.6 )\times 10^{66} \text{ cm}^{3}$, for the eastern and the western lobe respectively. Using the value of the pressure in the middle of the lobes from the southern profile found earlier, we estimated an enthalpy of $(1.1 \pm 0.5 )\times 10^{56}$ erg for the eastern lobe and $(3 \pm 1 )\times 10^{56}$ erg for the western lobe.

Considering that the lobes have risen buoyantly at a terminal velocity of $\sim 0.4 c_s$ \citep{churazov_evolution_2001} to the position of the center of each lobe, both traveling a distance of $\sim 1'$ (5 kpc), they are $\sim 20$ Myr old. 
Here we assume that the sound speed in the ICM, $c_{s}$, is $\sqrt{\gamma kT / (\mu m_{H})}$, where $\gamma$ is the ratio of specific heat capacities set to $5/3$ and $T$ is the the gas temperature in the middle of each lobe.
The terminal velocity of both lobes is then around $220 \text{ km s}^{-1}$. This gives an approximate outburst power of $(1.8 \pm 0.9)\times 10^{41}$ erg s$^{-1}$ for the eastern lobe and $(6 \pm 3)\times 10^{41}$ erg s$^{-1}$ for the western lobe.
These results are summarized in Table \ref{tab:cavities_energetic}. 


 { \renewcommand{\arraystretch}{1.2}
\begin{table*}
\centering
 \caption{Properties of the eastern and western cavities.}
 \label{tab:cavities_energetic}
 \begin{tabular}{p{5.5cm} | p{3.5cm}  p{3.5cm}  }
     &   Eastern lobe    &    Western lobe     \\
\hline
\rr Volume of the lobe                          & $(8 \pm 4 )\times 10^{65} \text{ cm}^{3}$ & $(1.5 \pm 0.6 )\times 10^{66} \text{ cm}^{3} $   \\
\rr Distance traveled by the lobe               & 4.6 kpc  & 3.5 kpc   \\
\rr Hydrogen density in the center of the lobe  & $0.009 \text{ cm}^{-3}$ & $0.013 \text{ cm}^{-3}$   \\
\rr Temperature in the center of the lobe       & 1.2 keV & 1.1 keV   \\
\rr Pressure in the center of the lobe          & $3.7\times 10^{-11} \text{ dyn cm}^{-2}$ &  $5.0\times 10^{-11} \text{ dyn cm}^{-2}$  \\
\rr Bubble enthalpy ($4PV$ per lobe)            & $(1.1\pm0.5)\times 10^{56}$ erg & $(3\pm1 )\times 10^{56}$ erg \\
\rr Sound speed in the center of the lobe        & $550 \text{ km s}^{-1}$ & $540 \text{ km s}^{-1}$  \\
\rr Age                                         & 20 Myr & 16 Myr \\
\rr Power                                       & $(1.8\pm0.9)\times 10^{41}$ erg s$^{-1}$  & $(6\pm3)\times 10^{41}$ erg s$^{-1}$  \\
\hline
 \end{tabular}
\end{table*}
}

 { \renewcommand{\arraystretch}{1.2}
\begin{table*}
\centering
 \caption{Energy injected in cavities compared to uplift energy.}
 \label{tab:uplift_energetics}
 \begin{tabular}{p{5.5cm} | p{3.5cm}  p{3.5cm} }
     & Eastern lobe & Western lobe      \\
  \hline
\rr Distance of the rim from the center & 7.9 kpc & 4.8 kpc \\
\rr Volume of the rim            &$(1.4\pm0.5)\times 10^{66} \text{ cm}^{3}$ & $(3\pm1)\times 10^{66} \text{ cm}^{3}$ \\
\rr Uplift mass                &$(9\pm3)\times 10^{6} \text{ M}_{\odot}$ & $(28\pm9)\times 10^{6} \text{ M}_{\odot}$  \\
\rr Uplift energy                   & $(1.1\pm 0.3)\times 10^{56}$ erg & $(3\pm 1)\times 10^{56}$ erg    \\
\rr Energy injected in the cavities & $(1.1\pm 0.5)\times 10^{56}$ erg & $(3\pm 1)\times 10^{56}$ erg  \\
\hline
 \end{tabular}
\end{table*}
}

\subsection{Uplift energy}
\label{Uplift energy}

Considering that the bright rims consist of cold gas probably dragged out from the center by buoyantly rising bubbles, we can estimate the energy required to lift this amount of gas from the center to the boundary of the lobe against gravity. We followed a procedure similar to \cite{reynolds_deep_2008} for A4059  and \cite{gitti_chandra_2011} for Hydra A, assuming an undisturbed ICM approximately isothermal in a hydrostatic configuration with density profile $\rho(r)$. The uplift energy is then given by:
\begin{eqnarray}
\Delta E = \dfrac{M_{cool} c_{s}^{2}}{\gamma} \ln \left( \dfrac{\rho_{i}}{\rho_{f}} \right)  \;,
\label{uplift_energy}
\end{eqnarray}
where $M_{cool}$ is the mass of the uplifted gas, $c_{s}$ is the sound speed in the ICM, $\gamma$ is the ratio of specific heat capacities, set to $5/3$, $\rho_{i}$ and $\rho_{f}$ are the initial and final densities of the undisturbed ICM. 

\begin{figure}
 \includegraphics[width=\columnwidth]{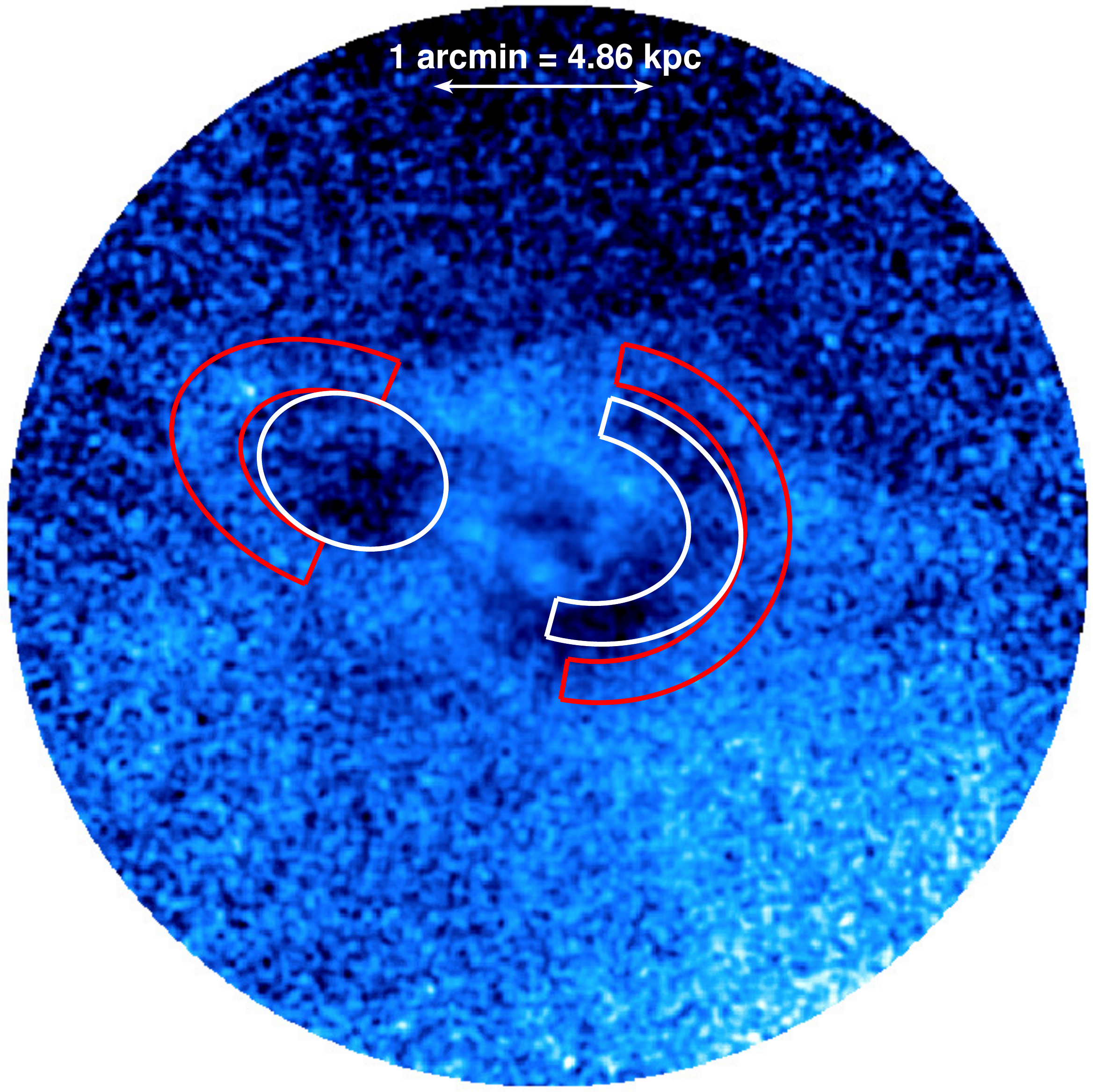}
 \caption{Fractional residual image in the 0.5-2 keV band, smoothed with a 2 pixel Gaussian function. We have chosen an ellipsoid shape (in white) for the eastern lobe volume with semi-major and minor axes of $0.44 \arcmin$ (2.13 kpc) and $0.34 \arcmin$ (1.64 kpc), respectively. For the western lobe, we have chosen half of an ellipsoidal shape (in white), with maximum semi-major and minor axes of $0.56 \arcmin$ (2.72 kpc) and $0.75 \arcmin$ (3.65 kpc) respectively and minimum semi-major and minor axes of of $0.38 \arcmin$ (1.85 kpc) and $0.51 \arcmin$ (2.48 kpc) respectively. The regions used to calculate the amount of gas uplifted in subsection \ref{Uplift energy} are in red.}
 \label{fig:ngc4472_merged_dmfilth_500-2000ev_unsharpmask_lobes_and_rims}
\end{figure}

We estimated the volume of the uplifted gas based on the shape of the rims visible in the fractional residual image (Fig.~ \ref{fig:ngc4472_merged_dmfilth_500-2000ev_unsharpmask_lobes_and_rims}). 
Again, the largest source of uncertainty in the estimation of the uplift energy comes from the unknown 3D shapes of the rims. 
Similarly to the calculation of the lobe volumes, we considered half-ellipsoidal shells, choosing a minimum and a maximum size for the rims, giving volumes of $(1.4\pm0.5)\times 10^{66} \text{ cm}^{3}$ and $(3\pm1)\times 10^{66} \text{ cm}^{3}$ for the eastern and the western lobes respectively.
From the density in the center of the rims obtained from the profile of Fig.~\ref{fig:plot_profiles} (the density in the center of each rim is $0.012 \text{ cm}^{-3}$ and $0.019 \text{ cm}^{-3}$), the estimated uplifted gas mass is $(9\pm3)\times 10^6 \text{ M}_{\odot}$ and $(28\pm9)\times 10^6 \text{ M}_{\odot}$. Using the temperature at the center of each rim ($1.1$ and $1.2$ keV, also from the previously calculated profile), the sound speed was calculated to be $540$ and $550 \text{ km s}^{-1}$ respectively. Finally, we estimated the uplift energy to be $\sim (1.1\pm0.3)\times 10^{56}$ erg for the eastern rim and $\sim (3\pm1)\times 10^{56}$ erg for the western rim. This analysis is summarized in Table \ref{tab:uplift_energetics}. 
In both cases, considering the uncertainties, the energy required to lift the gas in the rims from the center to the boundary of the lobes against gravity is comparable to the energy injected by the jets into the lobes.


We note that the rims appear against projected emission from the surrounding atmosphere and that all the gas in the rims might not be entirely uplifted gas. Therefore, part of the emission in the rims might not be coming from the uplifted gas and the uplift energy calculated could be lower.
Furthermore, the simplifying assumption that all of the gas in the shell rose buoyantly from the galaxy center is probably an oversimplification for several reasons, and significantly overestimates the energy in the uplift.  The total mass of the gas in the shell is quite large, and it is not plausible that it came from a small region at the galaxy center.  If the gas in the shell rose adiabatically, its entropy will be indicative of the height from which it was lifted.  Using the measured temperature and density of the gas shell around the Eastern lobe, we find it has an entropy of $kTn_{e}^{-2/3} \sim$26 keV cm$^{2}$. Estimating the entropy profile of the group gas from the temperature and density profiles in Figure~4, the entropy of the gas in the rim roughly equals the entropy of the group gas at a height of $\sim$4 kpc from the nucleus.  The energy required to uplift the gas from this height to its
present position is $\sim$20-25\% of the energy required to uplift the gas from the center of the galaxy.  We note that this number is still large compared to the equivalent heat input of the entropy rise due to weak shocks in mildly supersonic outbursts.  Such mildly supersonic outflows are, in themselves, considered to be sufficient to offset the radiative cooling of cool core clusters \citep{mcnamara_heating_2007}.  If a large fraction of this uplift energy is thermalized via hydrodynamic instabilities or thermal conduction, it could be an important or even dominant source of heating of cool core clusters and groups.  A detailed discussion of the dynamics of buoyancy and its impact on systems similar to NGC 4472 will be presented in a future publication (Su {\it et al.}, in preparation), but it is clear that this buoyancy and uplift may be an energetically important process in the thermodynamic lifecycle of cluster and group cores.

\section{Abundance maps and profiles}
\label{Abundance maps and profiles}

With the high number of counts in the Chandra observations of NGC 4472, its low temperature and relatively high metallicity, it is an ideal test case to study the abundance distribution in groups of galaxies. 
In particular, we have explored two aspects of the metallicity distribution in groups and clusters.
First, we looked for potential alignments of abundance structures with the radio outbursts as seen in several clusters e.g.; in Virgo \citep{simionescu_metal-rich_2008}, Hydra A \citep{gitti_chandra_2011}, Perseus \citep{sanders_non-thermal_2005}, RBS 797 \citep{doria_chandra_2012} and others \citep{kirkpatrick_anisotropic_2011,kirkpatrick_hot_2015} and at least one group (e.g. in HCG 62 \citealt{rafferty_deep_2013}).
Second, we investigated the presence of a central abundance decrement, a feature reported in many groups e.g. in NGC~5813 \citep{randall_very_2015} and clusters e.g. in Perseus \citep{sanders_deeper_2007}, Centaurus \citep{sanders_spatially_2002,panagoulia_searching_2013,sanders_very_2016}, Ophiuchus \citep{million_ram-pressure_2010}, Abell 2199 \citep{johnstone_chandra_2002} and others, \citep{panagoulia_volume-limited_2015}.

\begin{figure*}
\gridline{\figdiff{ds9_merged_500-2000ev_box3_smooth3_profile_test3}{0.315\textwidth}{(a) Sectors used for the profiles in Fig.~\ref{fig:ab_vapec_profiles}.}{20pt}\label{fig:ds9_merged_500-2000ev_box3_smooth3_profile_test2}
          \fig{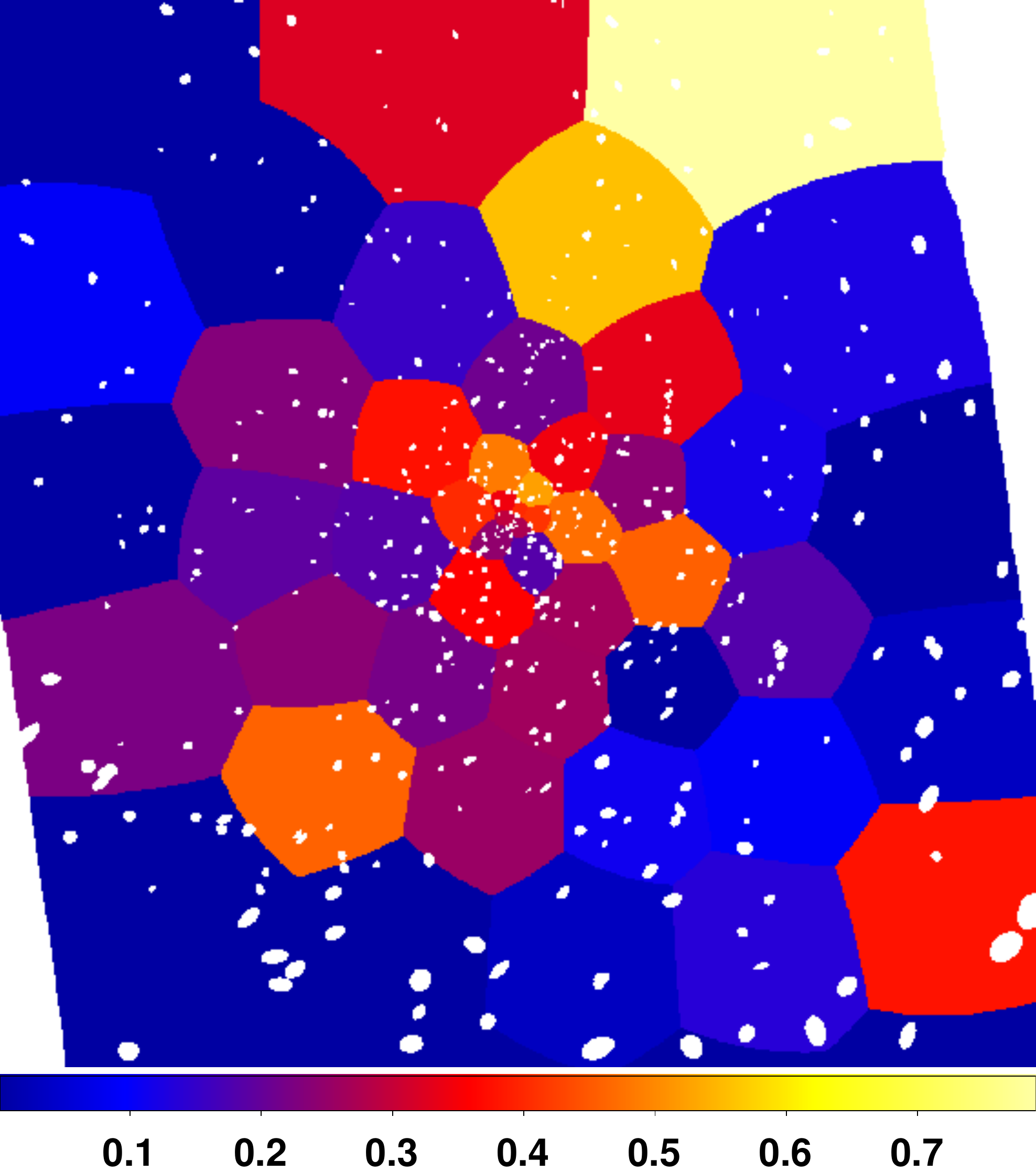}{0.315\textwidth}{(b) O abundance map.}\label{O_map}
          \fig{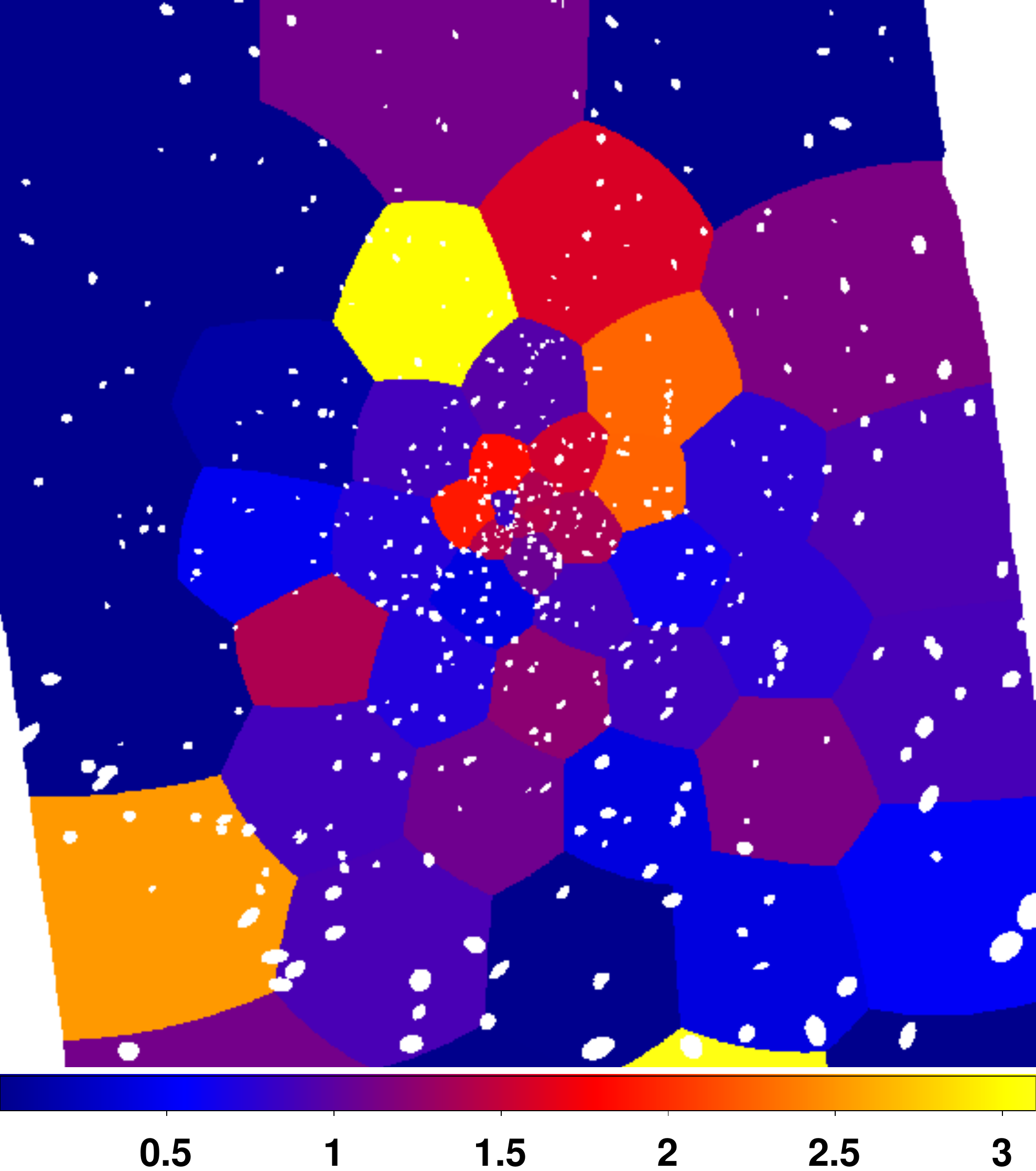}{0.315\textwidth}{(c) Ne abundance map.}\label{Ne_map}}       
\gridline{\fig{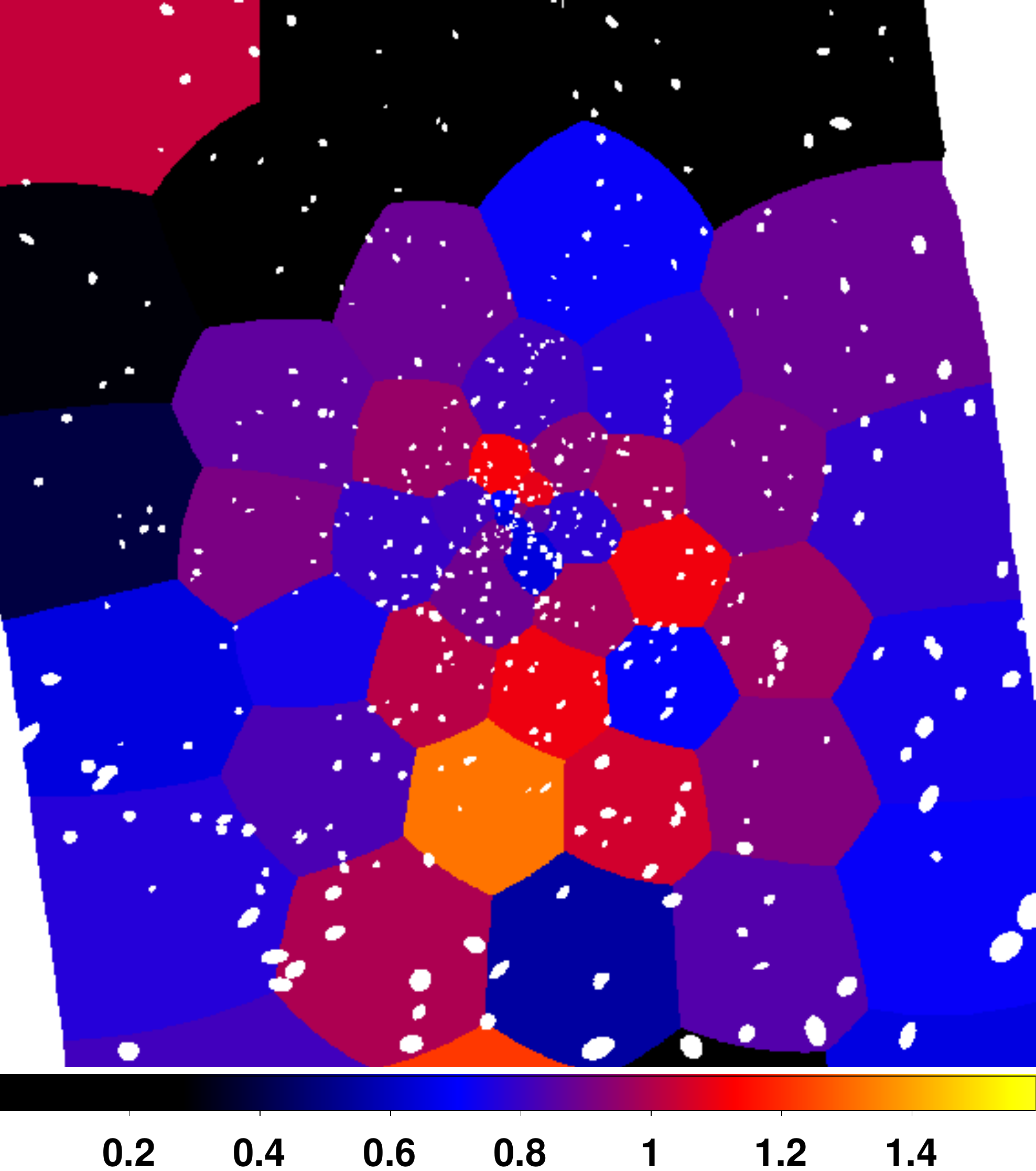}{0.315\textwidth}{(d) Mg abundance map.}\label{Mg_map}
          \fig{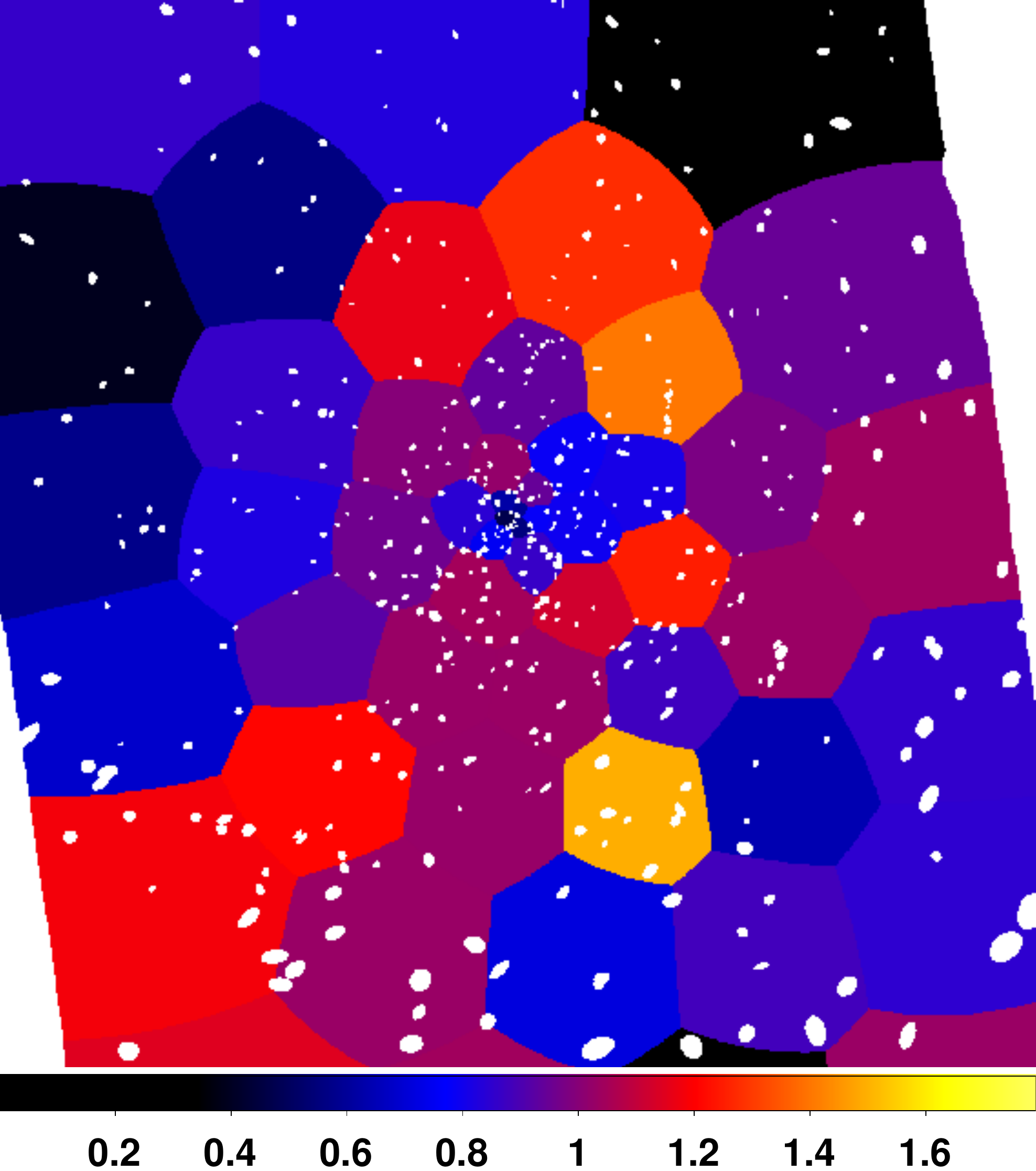}{0.315\textwidth}{(e) Si abundance map.}\label{Si_map}
          \fig{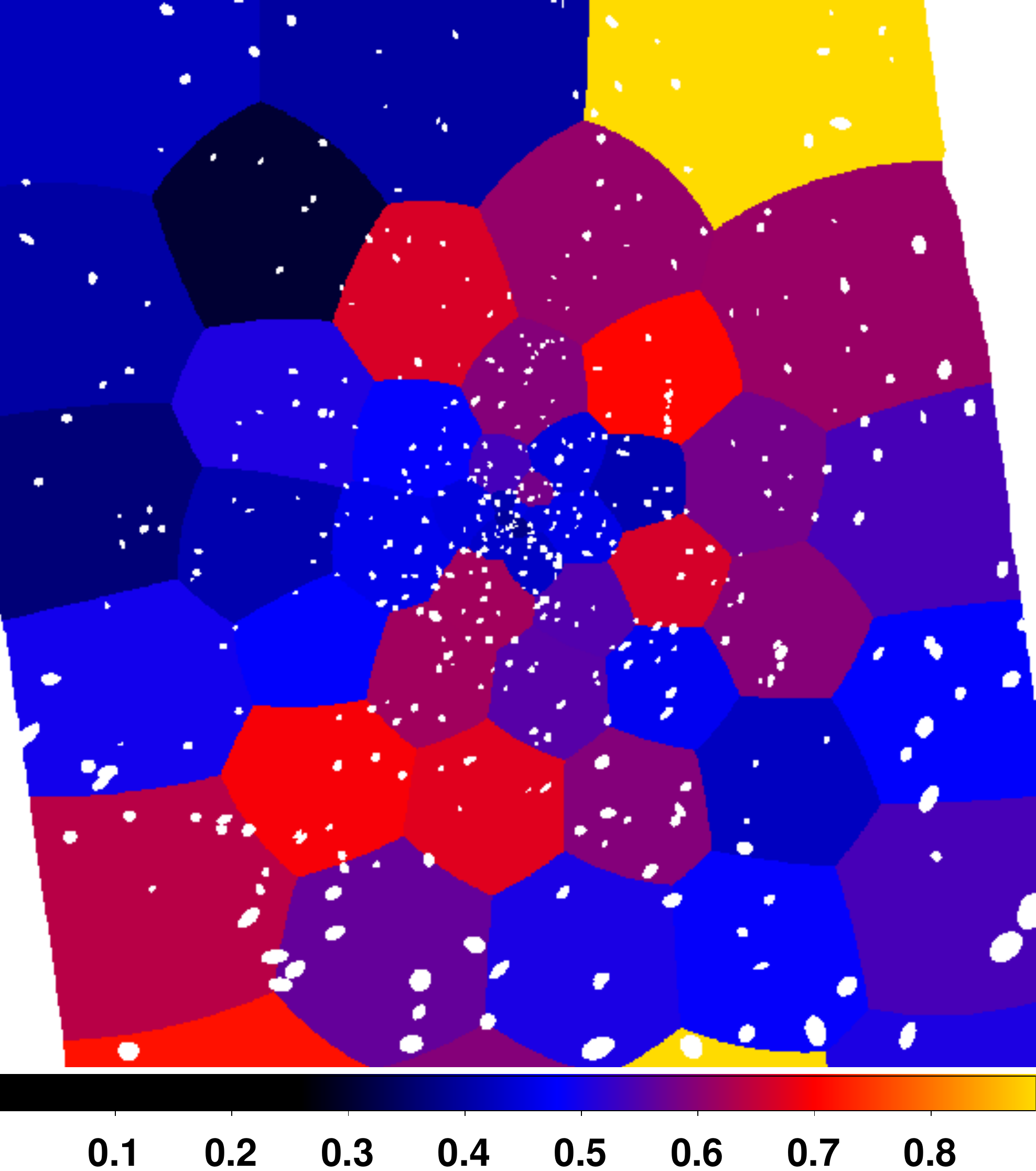}{0.315\textwidth}{(f) Fe abundance map.}\label{Fe_map}}
\caption{Merged image and abundance maps, all having the same size and scale, indicated in the lower-left corner of a). a) Merged image of the observations in the soft band (0.5-2 keV) smoothed with a 3 pixel Gaussian function showing the sectors used to extract the profiles in Fig.~\ref{fig:ab_vapec_profiles}. The pink regions enclose the region where the radio emission is found and the green sectors delineate the undisturbed ICM. b)-f) Abundance maps from the regions produced with a WVT algorithm for a signal to noise of 100. The scale bars give the abundance of each element with respect to solar units. The uncertainties on the abundances vary roughly from $35\%$ for O and Ne, to $15\%$ for Mg, $11\%$ for Si and $10\%$ for Fe.}
\label{fig:metallicity_map_corz_vapec_modif}
\end{figure*}

To look for a correspondance between the distribution of various metals in the ICM and the direction of the outflows, we produced abundance maps following a similar technique as in \cite{kirkpatrick_hot_2015}. 
We used a weighted Voronoi tessellation (WVT) algorithm (\citealt{cappellari_adaptive_2003}, \citealt{diehl_adaptive_2006}), choosing a target signal to noise of 100 per binned region. The merged image in the broad band (0.5 to 7 keV)  with point sources removed and the corresponding Poisson noise were given as inputs for the algorithm. We used again a \textsc{phabs$\ast$vapec} model to fit the data of each bin as described in section \ref{Enhanced X-ray emission rims}. The result is shown in Fig.~\ref{fig:metallicity_map_corz_vapec_modif}. 
The model fits reasonably well the data for most of the bins since the reduced chi-squared value was low (mostly below 2).
However, the residuals of some bins show that data goes above and under the model, suggesting the presence of multitemperature gas.

    \begin{figure*} 
    \includegraphics[width=\columnwidth]{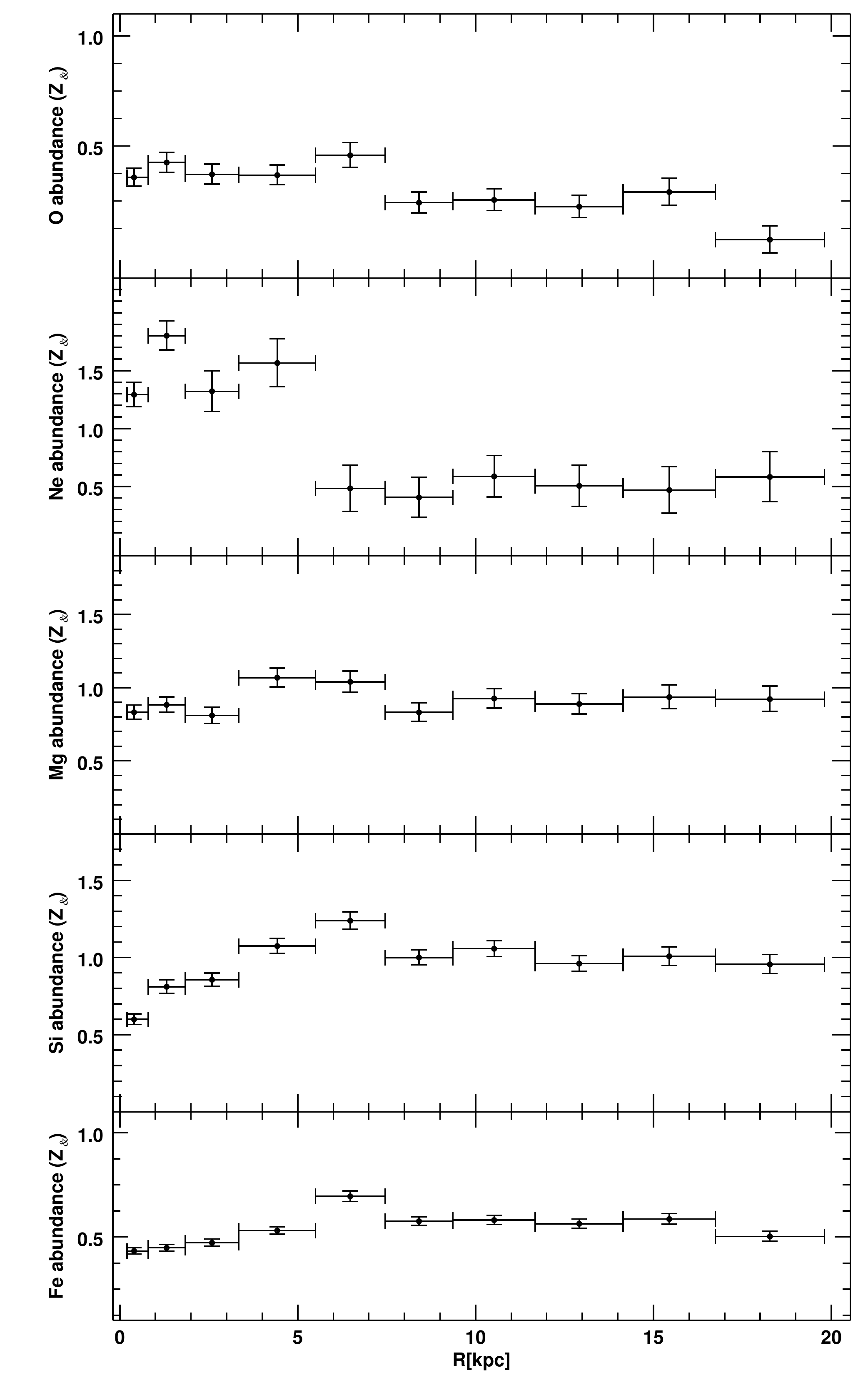}
    \includegraphics[width=\columnwidth]{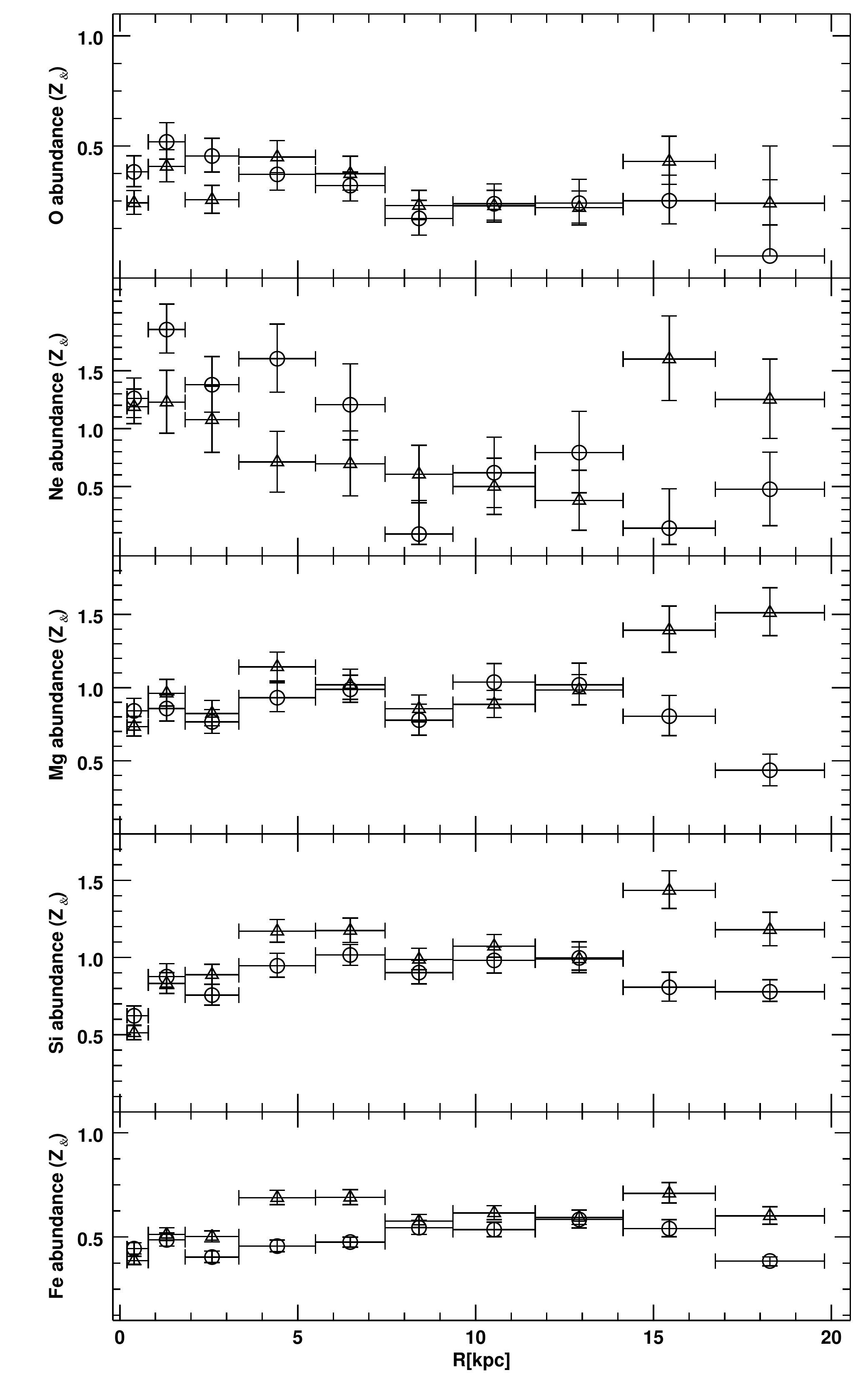}
\caption{\textbf{Left:} Abundance profiles of complete annuli ($360\,^{\circ}$) with the same width as the sectors from Fig.~\ref{fig:ds9_merged_500-2000ev_box3_smooth3_profile_test2}. \textbf{Right:} Abundance profiles of the on-jet sectors (circles) and the off-jet sectors (triangles) from the regions shown in Fig.~\ref{fig:ds9_merged_500-2000ev_box3_smooth3_profile_test2}.}
\label{fig:ab_vapec_profiles}
   \end{figure*}

We did not correct for projection effects because of the complicated temperature structure. We expect that removing the higher metallicity gas of the outer layer from the inner regions will increase the magnitude of the abundance drop.

As in \cite{kirkpatrick_hot_2015}, we also compared the metallicity profiles extracted from on-jet and off-jet sectors, respectively in the direction of the radio emission and in the orthogonal direction. The chosen semi-annular bins are shown on Fig.~\ref{fig:metallicity_map_corz_vapec_modif} a) and the profiles are presented in Fig.~\ref{fig:ab_vapec_profiles}-right. 
Each of these bins has a signal to noise ratio high enough that the uncertainties on the temperature and abundance are around $1\%$ and  $10\%$, respectively.
The same single temperature \textsc{phabs$\ast$vapec} \textsc{xspec} model as for the maps was used. These  semi-annular profiles can be compared with abundance profiles of complete annuli with the same width (Fig.~\ref{fig:ab_vapec_profiles}- left).

\section{Discussion}\label{Discussion}

We have conducted an X-ray analysis of the hot gas atmosphere of the nearby early-type galaxy NGC 4472, focussing on the central $\sim 20 \text{ kpc}$.
We have measured the temperature, pressure, and elemental abundance of the ICM along different regions. These measurements have allowed us to estimate of the outburst energy and power of the AGN.
We have determined that the rims of enhanced X-ray emission beyond the inner cavity pair cannot be AGN-driven shocks but are rather more consistent with cooler low entropy material being dragged out from the galaxy center. 
We have also looked for metallicity structures; alignment with the cavity system and central abundance decrement.
 
At this point it is relevant to compare the case of NGC 4472 to other groups and clusters. 
Indeed, other nearby galaxy groups also show signatures of significant impact from AGN outbursts e.g. AGN-driven shocks around the bubbles in NGC 4636 \citep{baldi_unusual_2009}, uplifted gas in NGC 5044 \citep{david_isotropic_2009,david_active-galactic-nucleus-driven_2011}, waves associated with the mildly supersonic bubble expansion in the Virgo elliptical galaxy M84 \citep{finoguenov_-depth_2008} or the presence of a sharp edge in NGC 507 associated with the transonic inflation of a bubble \citep{kraft_unusual_2004}.
In the following subsections, we discuss various implications of these results and compare them to other groups and clusters.

\subsection{Gas uplift}

Gas uplifts driven by AGN outbursts have been observed in several groups (e.g. NGC 5044, \citealt{david_isotropic_2009,david_active-galactic-nucleus-driven_2011}; HCG 62 \citealt{rafferty_deep_2013}) and clusters (e.g. M87, \citealt{forman_filaments_2007}; A4059, \citealt{reynolds_deep_2008}; Hydra A \citealt{gitti_chandra_2011}; \citealt{kirkpatrick_hot_2015} and references therein) at X-ray wavelengths but more recently also in CO measurements (e.g. \cite{russell_alma_2016}).

We can compare NGC 4472 with the sample of 29 galaxy groups and clusters with \textit{Chandra} observations hosting brightest cluster galaxies (BCGs) with X-ray cavities from \cite{kirkpatrick_hot_2015}. The authors found the presence of metal-rich plasma projected preferentially along cavities for 16 clusters. This material has been lifted out by rising cavities up to 20 to several hundreds of kiloparsecs. They identified the maximum projected distance of the uplifted gas from the center as the ``iron radius". As in \cite{kirkpatrick_anisotropic_2011}, they found a power-law relationship between jet power and the iron radius:
\begin{eqnarray}
 R_{Fe} = \left( 62 \pm 26 \right)  \times P_{jet}^{\left( 0.45 \pm 0.06\right) } \text{(kpc)} \,
 \label{eq:p_jet}
\end{eqnarray}
 where $P_{jet}$ is in units of $10^{44} \text{ erg s}^{-1}$, 
 and between the cavity enthalpy and the iron radius:
\begin{eqnarray}
 R_{Fe} = \left( 57 \pm 30 \right)  \times E_{cav}^{\left( 0.33 \pm 0.08\right) } \text{(kpc)} \,
 \label{eq:e_cav}
\end{eqnarray}
 where $E_{cav}$ is in units of $10^{59} \text{ erg}$.
Using cylindrical volumes with the same radius as the cavity and the length of the iron radius, they calculated the uplift energy given by equation \ref{uplift_energy}. They found that the fraction of uplift energy to the total cavity energy varies from a few percent to $50\%$, with an average of $14\%$.
We note that the uplift energies estimated in our case for the eastern lobe and the western lobe are comparable to the outburst energy, on the same scale as to the galaxy clusters and groups in \cite{kirkpatrick_hot_2015} but potentially larger. This emphasis that this uplift energy could be an important source of heating in groups and clusters. 
The equivalent of the iron radius in the case of NGC 4472 will be presented in section \ref{Correlation of the metallicity distribution with the cavity system}.

\subsection{Metallicity distribution in NGC 4472}

\subsubsection{The central abundance drop}

The presence of the large elliptical galaxy NGC 4472 at the center of a group would normally inject metals in its immediate neighborhood, inducing a peaked metallicity distribution. Instead, we note that, unlike the maps in \cite{kirkpatrick_hot_2015}, the central region seems to have a lower metallicity than the surrounding gas. 
We observe from Fig.~\ref{fig:ab_vapec_profiles}-left the presence of a peak for all abundances at a distance of $\sim 6$ kpc from the center as well as a drop from the peak value varying from 18 to $52 \%$ toward the center. This drop is $1\sigma$ in the case of the O and Ne profile,  $2\sigma$ for the Mg, $7\sigma$ for the Si and $6 \sigma$ for the Fe.
Interestingly, all abundance profiles show a flat behaviour beyond the core region from $\sim 8$ to $\sim 20$ kpc.

As mentioned earlier, central abundance drops have been found in other groups and clusters (\citealt{churazov_xmm-newton_2003,churazov_xmm-newton_2004}; \citealt{panagoulia_volume-limited_2015} and references therein).
\cite{panagoulia_volume-limited_2015} found 8 certain cases and 6 possible central abundance drops out of a sample of 65 groups and clusters with X-ray cavities. The radial iron abundance profiles of those sources had a central drop followed by a peak and a gradual decrease.
The authors suggest that the lack of iron in the core is due to the fact that iron depletes on to grains in cold dust, which are dragged outwards with optical filaments by the buoyantly rising bubbles. These grains are eventually destroyed at a certain distance from the center, resulting in the observed peak in the Fe profile.  
We consider this interpretation as a potential explanation for the observed dip in our abundance profiles.
The high number counts in the Chandra observations of NGC 4472 has allowed us to produce profiles not just for Fe but for several other elements. Our results suggest that other elements must also be dragged out (possibly from depletion on to dust grains).

A factor influencing the strength of the emission lines that needs to be considered in this analysis is resonance scattering. 
As resonance lines are produced by transitions involving the ground state of an ion, they occur more frequently than other transitions. For an ion column density high enough, this radiation can be scattered away from our line of sight. Considering that several X-ray emission lines are resonance lines, if the gas is optically thick at the energies of strong resonance lines, it will tend to drive the measured abundance to a lower value in the center of the galaxy group or cluster and overestimate it in the outskirts.
This effect was studied by \cite{sanders_resonance_2006}, where a spectral model accounting for resonance scattering was fitted to \textit{Chandra} observations of the Centaurus and Abell 2199, two clusters where a central abundance drop was observed. The authors concluded that the effect of resonance scattering on the metallicities was less than $10\%$, and could not explain the central dip.
\cite{panagoulia_volume-limited_2015} also discuss the role of resonance scattering and conclude that, considering the spectral resolution of \textit{Chandra}, the effect on their abundance profiles is negligible.
Based on the scattering calculation of \cite{mathews_spatial_2001,sanders_resonance_2006}, we found the strongest lines-center optical depths in the radial direction by integrating the scattering coefficients using the density profile of Fig.~\ref{fig:plot_profiles}. The largest optical depth obtained is 0.1 for the \ion{O}{8} line.
Since the abundance drop from the peak value varies from 18 to $52 \%$, as calculated with the values shown in Fig.~\ref{fig:ab_vapec_profiles}-left, we therefore conclude that the effect of resonance scattering is negligible. 


Since the fit of a 1-T model to a multitemperature gas can underestimate the abundance values found (i.e. the iron bias, \citealt{buote_x-ray_2000,Buote_xmm-newton_2003}), we also tested a \textsc{phabs$\ast$(vapec+vapec)} model to fit the data in complete ($360\,^{\circ}$) annular regions.
The second temperature component was fixed to the 1-T component (which was allowed to vary freely) plus 0.4 keV. This value was chosen after several tests, this one giving the best reduced chi-squared in general.
The statistical goodness of the fit was not significantly improved by the addition of a second temperature component. The reduced chi-squared values of each fit were only a few percent lower, except in the innermost bin, where the 1T model fit was better. 
The presence of a second temperature component was significant from the second to the last bins ($\sim 1\text{ kpc}<r<20\text{ kpc}$). In the first bin, the \textsc{xspec} norm related to the second temperature component was only 0.1\% of the norm of the first component.
Furthermore, the 2-T model resulted, in general, into larger errors bars for the different fitted parameters: 10\% to 50\% larger errors in abundances, up to $\sim 5$ times larger in temperature and up to $\sim 2$ times larger for the norm.
Finally, the profiles obtained with the 2T model still show the presence of a central drop for all abundances, varying from 24 to $58 \%$ of abundance drop from the peak value. 
We therefore conclude that these central drops persist with the fit of a multitemperature model to the annular regions.

\subsubsection{Correlation of the metallicity distribution with the cavity system} \label{Correlation of the metallicity distribution with the cavity system}


Inspection of the abundance maps in Fig.~\ref{fig:metallicity_map_corz_vapec_modif} b)-f) shows that any correlation between the metallicity distribution and the cavity system orientation is weak. 
For most abundance profiles in Fig.~\ref{fig:ab_vapec_profiles}- right, the values of both regions agree within $1\sigma$, implying a relatively homogeneous distribution of metals for a given radius.
However, surprisingly, the Fe abundance map shows some structure, indicating a slightly lower abundance in the direction of the jets, particularly to the east.
One possible explanation for this result comes from the previous observation that the central bins of the metallicity map have lower metallicity: in this case, we expect that the regions along the jets, where the gas of the BGG is deposited, has a similar abundance, which is lower than the surrounding undisturbed gas. 

We can define an ``iron radius" in a similar way as in \cite{kirkpatrick_hot_2015}: from the iron profile in Fig.~\ref{fig:metallicity_map_corz_vapec_modif}, counting bins outwards until the bins' error bars do not overlap and beyond which the on-jet regions has higher metallicity or the error bars overlap. According to this adapted definition, we find an iron radius of $6 \pm 1$ kpc. This value is consistent with the values predicted by the power-laws in Eq.~\ref{eq:p_jet} and \ref{eq:e_cav}, giving respectively $7\pm3$ kpc and $9\pm5$ kpc for the total jet power and cavity enthalpy of the two lobes system. However, we note that the uncertainties are quite large.




\section{Conclusions}\label{Conclusion}

In conclusion, we presented the analysis of almost 400 ks of ACIS-S \textit{Chandra} observations of the elliptical galaxy NGC 4472, consisting of 300 ks of new data. In summary, we conclude the following.

\begin{itemize}[itemsep=1pt, topsep=5pt, partopsep=0pt, leftmargin=0.5cm]
\item This work has enabled the detailed characterization of the two X-ray cavities of about 5 kpc diameter found in NGC 4472, filled by relativistic plasma seen at 1.4 GHz and surrounded by rims of enhanced X-ray emission.
\item The temperature profiles across the eastern and western rims have revealed that the gas in those rims is colder than the exterior gas, eliminating the possibility of AGN-driven shocks. We therefore argue that these rims are probably made of cold gas being dragged out by the two buoyantly rising bubbles.
\item We estimate that the energy required to lift this gas is comparable to the total outburst energy, both being roughly $10^{56}$ erg. If a large fraction of this uplift energy is thermalized via hydrodynamic instabilities or thermal conduction, this buoyancy and uplift may be an energetically important process in the thermodynamic lifecycle of cluster and group cores.
\item We found a central drop in the metallicity distribution of NGC 4472. In the case of the iron abundance, the central region and the sector in the direction of the cavity system both have a lower abundance than the surrounding ICM. If the central iron drop is real, then we suggest that the uplifted gas would also have a lower iron abundance. 
\end{itemize}


\acknowledgments
MLGM is supported by NASA grants GO1-12160X and NAS8-03060, as well as by NSERC through the Postgraduate Scholarships-Doctoral Program (PGS D) and Universit\'e de Montr\'eal physics department. MLGM thanks H. Intema for useful discussion on radio data.
JHL is supported by NSERC through the discovery grant and Canada Research Chair programs, as well as FRQNT.



\bibliographystyle{apj}
\bibliography{biblio_ngc4472}





\end{document}